\documentclass[12pt]{iopart}
\usepackage[latin1]{inputenc}
\usepackage[T1]{fontenc}
\usepackage[english]{babel}
\begin{document}
\title[Isotropisation of Generalized Scalar-Tensor theory plus a massive scalar field...]{Isotropisation of Generalized Scalar-Tensor theory plus a massive scalar field in the Bianchi type $I$ model}
\author{St\'{e}phane Fay}
\address{14 rue de l'Etoile\\
75017 Paris\\
France\footnote[2]{Steph.Fay@Wanadoo.fr}}
\date{\today}
\begin{abstract}
In this paper we study the isotropisation of a Generalized Scalar-Tensor theory with a massive scalar field. We find it depends on a condition on the Brans-Dicke coupling function and the potential and show that asymptotically the metric functions always tend toward a power or exponential law of the proper time. These results generalise and unify these of De Sitter in the case of a cosmological constant and of Cooley and Kitada in the case of an exponential potential.
\\
\\
\\
Published in Classical and Quantum Gravity copyright 2001 IOP Publishing Ltd\\
Classical and Quantum Gravity, Vol 18, 15, 2001.\\
http://www.iop.org
\end{abstract}
\maketitle
%------------------------------------------------------------------------------------------------------------------------------------------------------------------------------------------%
\section{Introduction}\label{s0}
In this paper we wish to study the conditions for isotropisation of the Generalized Scalar-Tensor theory plus a massive scalar field in the Bianchi type $I$ model. They are many reasons to be interesting by this model\\
First of all General Relativity is a good description for the weak gravitational fields (solar system tests) as for strong ones (binary pulsar) although deviating are expected for early times or in extreme cases such as black hole. Then, it is interesting to consider a Lagrangian whose "geometric" part looks like General Relativity. Moreover particle physics progress and the idea of inflation in the eighties show that scalar fields could be essential components of a gravitational theory. In this work we will consider a massive one. It could be justified by observations of type $IA$ supernovae \cite{Per99,Rie98} which seem to demonstrate that the dynamical behaviour of our Universe is accelerated. Most of time this is interpreted as the presence of a cosmological constant in the field equations although other explanations can be advanced such as this of a non-perfect fluid with quintessential matter \cite{ChiJakPav00}. The Boomerang experiment \cite{Lan00} indicates that it could represent the dominant energy of our Universe. However, the present value of this constant is in disagreement with this predicted by particle physics at early times. One way to solve this problem is to consider a varying potential $U$, i.e. a massive scalar field. We will also describe the coupling of the scalar field $\phi$ with the metric functions by a coupling function $\omega(\phi)$ generalising the Brans-Dicke coupling constant \cite{BraDic61}. This type of coupling is issued from particle physics theories whose Lagrangian at low energy could take the form of a scalar tensor theory.\\
Geometrically, our present Universe seems well described by the isotropic and homogeneous cosmological models, i.e. the FLRW models. However the observations, as instance from Boomerang \cite{Lan00}, show slight anisotropies in the cosmological microwave background, which could take origin at early times. Moreover, if the Universe had always been perfectly isotropic and homogeneous, it would be difficult to explain the large-scale structures we observe.  Hence, it is interesting to consider an anisotropic Universe described by the Bianchi models. There are 9 ones but the most studied are the Bianchi type $I$, $V$, $VII_0$, $VII_h$ and $IX$ which are able to isotropise toward an FLRW model \cite{ColHaw73}. We will consider the Bianchi type $I$ model which can tend toward a flat FLRW one and is a good candidate from the inflation theory point of view. Of course the Bianchi models are not a definitive geometrical description of the Universe which should probably be inhomogeneous. But such models allow studying the necessary conditions for its isotropisation.\\
Our goal will be to look for the necessary conditions depending on the potential and the Brans-Dicke coupling function for Universe isotropisation at late times. We will then derive the asymptotical dynamical behaviour of the metric functions and the condition for the presence of inflation.\\
Technically, we will use the ADM Hamiltonian formalism \cite{Nar72,Mis62} allowing to write the field equations as a first order system and then the dynamical systems theory as described in \cite{WaiEll97} and suggested in \cite{Fay00A} to study them. We have not found any paper in the literature where these two methods are applied to equations system with 2 arbitrary functions. Due to this indeterminacy  all the equilibrium points of the system can not be studied. However this problem can be overcomen for the subset of the phase space where lie the isotropic states of the Universe and which is of interest for us.\\
This paper is organised as follow. In the second section we calculate the Hamiltonian field equations of the Generalized Scalar-Tensor theory plus a massive scalar field and rewrite them with new normalised variables. In the third section, we study the subset of the phase space corresponding to isotropy. We discuss about physical meaning of the mathematical results thus obtained in the fourth section.
%------------------------------------------------------------------------------------------------------------------------------------------------------------------------------------------%
\section{Field equations}\label{s1}
The Lagrangian of the Generalized Scalar-Tensor theory plus a massive scalar field is written:
%----------------------------------EQUATION----------------------------------------%
\begin{equation} \label{1}
S=(16\pi)^{-1}\int \left[R-(3/2+\omega(\phi))\phi^{,\mu}\phi_{,\mu}\phi^2 -U(\phi)\right]\sqrt{-g}d^4 x
\end{equation}
with $\phi$ the scalar field, $\omega$ the coupling function between the scalar field and the metric, $U$ the potential. We will use the following form of the metric: 
\begin{equation} 
ds^2 = -(N^2 -N_i N^i )d\Omega^2 + 2N_i d\Omega\omega^i + R_0 ^2 g_{ij}\omega^i \omega^j 
\end{equation}
the $\omega^i$ being the 1-forms defining the Bianchi type $I$ homogeneous space. The $g_{ij}$ are the metric functions, $N$ and $N_i$ are respectively the lapse and shifts functions. Using the methods described in \cite{Nar72} and \cite{MatRyaTot73}, we find that the action can be rewritten in the following way: 
%-----------------------EQUATION-------------------------------------%
\begin{equation}\label{1a}
S=(16\pi)^{-1}\int(\Pi^{ij}\frac{\partial{g_{ij}}}{\partial{t}}+\Pi^{\phi}\frac{\partial{\phi}}{\partial{t}}-NC^0-N_iC^i)d^4x
\end{equation}
The $\Pi^{ij}$ and $\Pi^{\phi}$ are respectively the conjugate momentum of the  metric functions and scalar field, the $N$ and $N_i$ play the role of Lagrange multipliers. The quantities $C_0$ and $C_i$ are respectively the super-Hamiltonian and supermomentum defined by:
%----------------------------------EQUATION----------------------------------------%
\begin{equation}
C^0 =-\sqrt{^{(3)}g}^{(3)}R-\frac{1}{\sqrt{^{(3)}g}}(\frac{1}{2}(\Pi^k _k )^2 -\Pi^{ij}\Pi_{ij})+\frac{1}{\sqrt{^{(3)}g}}\frac{\Pi_\phi ^2 \phi^2 }{6+4\omega}+\sqrt{^{(3)}g}U(\phi)
\end{equation}
%----------------------------------EQUATION----------------------------------------%
\begin{equation}
C^i =\Pi^{ij}_{\mid j}
\end{equation}
where the "$^{(3)}$" hold for the quantities calculated on the 3-space and the "$\mid$" for the covariant derivative in the 3-space. When we vary the action with respect to the Lagrange multipliers we find the constraints $C^0=C^i=0$.\\
We rewrite the metric functions as $g_{ij}=e^{-2\Omega+2\beta_{ij}}$ and we use the Misner parameterisation \cite{Mis62}: 
%-----------------------EQUATION-------------------------------------
\begin{equation}
p_k^i=2\pi\Pi_k^i-\frac{2}{3}\pi\delta_k^i\Pi_l^l
\end{equation}
%-----------------------EQUATION-------------------------------------%
\begin{equation}
6p_{ij}=diag(p_++\sqrt{3}p_-,p_+-\sqrt{3}p_-,-2p_+)
\end{equation}
%-----------------------EQUATION-------------------------------------%
\begin{equation}
\beta_{ij}=diag(\beta_++\sqrt{3}\beta_-,\beta_+-\sqrt{3}\beta_-,-2\beta_+)
\end{equation}
Then, the action (\ref{1a}) is written as: 
%-----------------------EQUATION-------------------------------------
\begin{equation}
S=\int p_+d\beta_++p_-d\beta_-+p_\phi d\phi-Hd\Omega
\end{equation}
with $p_\phi=\pi\Pi_\phi$ and $H=2\pi\Pi_k^k$. Finally, from the constraint $C^0=0$, we get the expression for the ADM Hamiltonian:
%----------------------------------EQUATION----------------------------------------%
\begin{equation} \label{4}
H^2 = p_+ ^2 +p_- ^2 +12\frac{p_\phi ^2 \phi^2}{3+2\omega}+24\pi^2 R_0 ^6 e^{-6\Omega}U
\end{equation}
from which we derive the Hamiltonian equations:
%----------------------------------EQUATION----------------------------------------%
\begin{equation} \label{5}
\dot{\beta}_ \pm = \frac{\partial H}{\partial p_ \pm}=\frac{p_\pm}{H}
\end{equation}
%----------------------------------EQUATION----------------------------------------%
\begin{equation} \label{6}
\dot{\phi}=\frac{\partial H}{\partial p_\phi}=\frac{12\phi^2 p_\phi }{(3+2\omega)H}
\end{equation}
%----------------------------------EQUATION----------------------------------------%
\begin{equation} \label{7}
\dot{p}_\pm=-\frac{\partial H}{\partial \beta_ \pm}=0
\end{equation}
%----------------------------------EQUATION----------------------------------------%
\begin{equation} \label{8}
\dot{p}_\phi=-\frac{\partial H}{\partial \phi}=-12\frac{\phi p_\phi ^2}{(3+2\omega)H}+12\frac{\omega_\phi \phi^2 p_\phi ^2 }{(3+2\omega)^2 H}-12\pi^2 R_0 ^6 \frac{e^{-6\Omega}U_\phi }{H}
\end{equation}
%----------------------------------EQUATION----------------------------------------%
\begin{equation} \label{9}
\dot{H}=\frac{dH}{d\Omega}=\frac{\partial H}{\partial \Omega}=-72\pi^2 R_0 ^6 \frac{e^{-6\Omega}U}{H}
\end{equation}
A dot means a derivative with respect to $\Omega$. We will choose $N^i=0$ and we calculate $N$ by writing that $\partial \sqrt{g}/\partial \Omega=-1/2\Pi^k_kN$ \cite{Nar72}. Then, it comes: 
%-------------------------EQUATION-------------------------------------%
\begin{equation} \label{9a}
N=\frac{12\pi R_0^3e^{-3\Omega}}{H}
\end{equation}
The relation between the $\Omega$ and $t$ times is then $dt=-Nd\Omega$. We want to rewrite some of these equations under the form of an autonomous first order system with normalised variables \cite{WaiEll97}. For this we will only consider the set of equations (\ref{6}), (\ref{8}) and (\ref{9}), the equations (\ref{7}) showing that the conjugate momentums of the variables describing the anisotropy are some constants. The constraint (\ref{4}) suggests the following set of normalised variables: 
%-------------------------EQUATION-------------------------------------%
\begin{equation} \label{10a}
x=H^{-1}
\end{equation}
%-------------------------EQUATION-------------------------------------%
\begin{equation} \label{10b}
y=\sqrt{e^{-6\Omega}U}H^{-1}
\end{equation}
%-------------------------EQUATION-------------------------------------%
\begin{equation} \label{10c}
z=p_{\phi}\phi(3+2\omega)^{-1/2}H^{-1}
\end{equation}
The first one depends on $H$, the second one on $H$ and $\phi$ and the third one on $H$, $\phi$ and $p_\phi$. Thus they are independent variables. $y$ and $z$ will be real if the functions $U$ and $3+2\omega$ are positives. Under these conditions, it follows that the potential will favour inflation and the coupling function, when $U=0$, will be such that the energy density of the scalar field is positive. Rewriting the constraint (\ref{4}) with the new variables, we get:
%-------------------------EQUATION-------------------------------------%
\begin{equation} \label{11}
p^2x^2+R^2y^2+12z^2=1
\end{equation}
The positive constants $p$ and $R^2$ are defined by $p^2=p_+^2+p_-^2$ and $R^2=24\pi^2R_0^6$. From this last equation we deduce that the variables $x$, $y$ and $z$ are bounded and belong to the following intervals:
%-------------------------EQUATION-------------------------------------%
\begin{eqnarray}
x\in\left[-p^{-1},p^{-1}\right]&&\\
y\in\left[-R,R\right]&&\\
z\in\left[-1/\sqrt{12},1/\sqrt{12}\right]&&\\\nonumber
\end{eqnarray}
The field equations (\ref{6}), (\ref{8}) and (\ref{9}), become (see appendix \ref{A1}): 
%-------------------------EQUATION-------------------------------------%
\begin{equation} \label{12}
\dot{x}=3R^2y^2x
\end{equation}
%-------------------------EQUATION-------------------------------------%
\begin{equation} \label{13}
\dot{y}=y(6\ell z+3R^2y^2-3)
\end{equation}
%-------------------------EQUATION-------------------------------------%
\begin{equation} \label{14}
\dot{z}=y^2R^2(3z-\frac{1}{2}\ell)
\end{equation}
with $\ell=\phi U_\phi U^{-1} (3+2\omega)^{-1/2}$. They can not be expressed only with $x$, $y$ and $z$ because we do not wish to specify the form of $\omega$ and $U$ which are arbitrary functions of the scalar field. However, we do not need to know the exact form of $\ell (x,y,z)$ since we are only interested by the asymptotical isotropisation of the Universe at late times. To reach our goal, it is sufficient to assume two types of asymptotical behaviours for $\ell$: ever it tends toward a constant or it diverges. This excludes any asymptotical chaotic behaviour for $\ell$ and is in accordance with much of the functions $\omega$ and $U$ studied in the literature. We will have to check if these behaviours are compatible with the isotropisation of the Universe at late times.\\
\\
In the next section, we examine the equations (\ref{12}-\ref{14}) from the dynamical systems theory point of view. Firstly, we look for monotonic functions and secondly, we study the presence of equilibrium points.
%------------------------------------------------------------------------------------------------------------------------------------------------------------------------------------------%
\section{Dynamical studies of the fields equations}\label{s2}
\emph{Monotonic functions}\\
\\
Lets examine the presence of monotonic functions. From the equation (\ref{12}), we deduce that $x$ is a monotonic function: when it is positive (negative), it increases (decreases). Since $x$ has a constant sign, it follows from (\ref{10b}) that it is the same for $y$. Notes also that in the plane $x=0$ with $\ell=cte$, $z$ is a monotonic and increasing function if $z>\ell/6$, decreasing otherwise. Thus there is no periodic or homoclinic orbit and then no chaotic behaviour.\\
If we look for the signs of the derivatives of the metric functions with respect to $\Omega$ depending on the position of a point $(x,y,z)$ in the phase space, we see that the sets of points such that they are constants are splat by planes defined by $x=cte$ since $dg_{ij}/d\Omega=-2e^{-2\Omega+2\beta_{ij}}(1-\dot{\beta}_{ij})$. As instance for $g_{11}$, it is defined by $x=(p_++\sqrt{3}p_-)^{-1}$ and the sign of its derivative above or below this plane depends on the value of the constant $p_++\sqrt{3}p_-$. For $g_{22}$, the plane is $x=(p_+-\sqrt{3}p_-)^{-1}$ and for $g_{33}$, $x=p_+^{-1}$. Since $x$ is a monotonic function with constant sign, we deduce that each metric functions can have one and only one extremum. From (\ref{9a}) and the relation between $\Omega$ and $t$, we remark it will be the same in the proper time $t$. This is in agreement with the results of \cite{Fay00A}\\
\\
\emph{Study of the isotropic equilibrium states}\\
\\
Now we study the equilibrium points. They are all defined by $(y,z)=(\pm(3-\ell^2)^{1/2}(3R^2)^{-1/2},\ell/6)$ and they will respect the constraint if $x=0$. We have shown in \cite{Fay00A} that the Universe isotropises in the proper time $t$ only when $\Omega\rightarrow -\infty$. This value of $\Omega$ indicates that they will be sources or sinks but not saddle points. Thus an equilibrium states will represent an isotropic one for the Universe if in the same time $\Omega$ diverges negatively. Moreover, since $x$ is a monotonic function of constant sign, we deduce from the relation (\ref{9a}) that the proper time is a monotonic and decreasing (increasing) function of $\Omega$ when the Hamiltonian is positive (negative). Hence $\Omega$ can be considered as a time variable and the equilibrium will take place at late times if $H>0$. In what follows, we will assume that $\ell$ asymptotically tends toward a constant or diverges.\\
\newline
\underline{First, we assume that $\ell$ is asymptotically a constant}. We can show in that case by integrating (\ref{13}-\ref{14}) that when $y=\pm(3-\ell^2)^{1/2}(3R^2)^{-1/2}$, $\Omega$ diverges. It follows that these two equilibrium points are compatible with the isotropisation of the Universe. $R^2$ being a positive constant, they will be real if $\ell^2<3$. We can not calculate their corresponding eigenvalues and thus knowing the signs of these last quantities because we do not now the expressions of the derivatives of $\ell$ with respect to $x$, $y$ and $z$. However, since we consider $\Omega$ as a time variable, they will be sinks if $H>0$ or sources if $H<0$ since they will respectively correspond to asymptotical late or early times.\\
To get the behaviour of the metric functions when we approach an isotropic state, we need a differential equation for $x$ when $\ell\rightarrow cte$. In this last case, the integration of the field equations (\ref{12}-\ref{14}) gives: 
%--------------------------------%
\begin{equation}
z=\left[\ell(1+6R^2z_0)-\sqrt{\ell^2(1+6R^2z_0)+18R^2z_0(R^2y^2-1)}\right](36R^2z_0)^{-1}
\end{equation}
By introducing this expression in the constraint equation and using (\ref{12}) to express $y$ as a function of $x$ and its derivative, we get a differential equation for $x$. Since when the Universe isotropises, $x$ and its derivative tend toward zero as $\Omega$ diverges, and keeping only the second order terms in $x$ and $\dot{x}$, we find that $x$ asymptotically behaves as $x_0exp\left[(3-\ell^2)\Omega\right]$ when it tends to vanish. Taking into account the divergence of $\Omega$, we see that our approximation will be justified if $\ell^2<3$, which is in accordance with our previous results. One could also recover this result by linearizing the equation (\ref{12}) but we find this demonstration more rigorous.\\
\underline{Now we examine the case for which $\ell$ diverges}. It implies that the equilibrium points are unbounded. However, since $y$ and $z$ are bounded, we deduce that an equilibrium state can not be reached when $\ell$ diverges.\\
\\
In the next section, we discuss about physical meaning of our results. 
%------------------------------------------------------------------------------------------------------------------------------------------------------------------------------------------%
\section{Discussion}\label{s3}
In this work, we have examined the conditions under which the Universe described by a Generalized Scalar-Tensor theory with a massive scalar field isotropises as well as the asymptotical behaviour of the metric functions by help of the Hamiltonian formalism and dynamical systems theory.\\
The set of points of the phase space corresponding to stable isotropic states for the Universe is such that the time coordinate $\Omega$ and the Hamiltonian diverges ($\Omega\rightarrow -\infty$ and $x\rightarrow 0$). Then, the functions $\beta_\pm$ describing the anisotropy asymptotically tend toward a constant. We have shown that when $\ell$ was asymptotically unbounded, an equilibrium state could not be reached. Thus, the isotropy of the Universe is not compatible with the divergence of the quantity $\phi U_\phi U^{-1} (3+2\omega)^{-1/2}$. However it arises when its value belongs to the range $\left[-\sqrt{3},\sqrt{3}\right]$. In this case, the plane $x=0$ contains two equilibrium points corresponding to an isotropic state for the Universe. They are late times attractors in the $t$ time if the Hamiltonian is a positive function. If $H$ is interpreted as an energy, it means that we assume a positive energy for the Universe, which is reasonable. If it is not the case, the isotropisation arises at early times. Moreover, we have shown that as long as $\ell^2<3$, the function $x(\Omega)$ asymptotically tended toward $x_0 exp\left[(3-\ell^2)\Omega\right]$. Using (\ref{9a}), we see that it corresponds to a power law of the proper times with the exponent $\ell^{-2}$ if $\ell$ does not tend toward a vanishing constant, or toward an exponential of $t$ otherwise. These two types of functions represent the only possible attractors when the Universe isotropises. All this can be summarised in the following important result:\\
\\
\emph{A necessary condition for isotropisation of the Generalized Scalar-Tensor theory plus a massive scalar field $\phi$, whatever the Brans-Dicke coupling function $\omega$ and the potential $U$ considered, will be that $\phi U_\phi U^{-1} (3+2\omega)^{-1/2}$ tends toward a constant $\ell$ with $\ell^2<3$. It arises at late times if the Hamiltonian is positive, at early times otherwise. If $\ell\not =0$ the metric functions tend toward $t^{\ell^{-2}}$. The Universe is expanding and will be inflationary if $\ell^{2}<1$. If $\ell=0$, the Universe tends toward a De Sitter model.
}\\
\\
Note, that the asymptotical behaviour of the metric functions when isotropisation arises does not depend on initial conditions whereas the epoch of isotropisation, i.e. late or early times, depends on the initial sign of the Hamiltonian. One element is missing in this result: the value of the scalar field when the isotropisation arises, i.e. when $\Omega\rightarrow -\infty$. Expressing $\dot{\phi}$ as a function of $z$ and $\phi$ (see appendix \ref{A1}), and taking $z$ as its value at the equilibrium, $\ell/6$,we get a differential equation for $\phi$. It does not describe the scalar field behaviour during the whole  Universe evolution, but asymptotically when $\Omega\rightarrow -\infty$ and the system approach equilibrium. This equation of the first order can be solved analytically or numerically. This additional result allows to calculate $\ell$ when isotropisation occurs and completes the main one above. It is written:\\
\\
\emph{The value of the scalar field when the Universe reaches an isotropic equilibrium state is the value of the function $\phi$ defined by $\dot{\phi}=2\phi^2U_\phi(3+2\omega)^{-1}U^{-1}$ when $\Omega\rightarrow -\infty$.}\\
\\
Lets examine the relations between these results and others quoted in the literature.\\

\emph{Firstly} they are in accordance with the "No Hair Theorem" which states that General Relativity with a cosmological constant isotropises toward a De Sitter model since in this case, $\ell=0$. It will be true for any form of potential and Brans-Dicke coupling function such that $\ell$ asymptotically vanishes when $\Omega\rightarrow -\infty$ which does not necessary implies that the potential tends toward a constant. As instance, it arises if the Brans-Dicke coupling function diverges faster than $\phi U_\phi U^{-1}$. This generalises the "No Hair Theorem" in the special case of the Bianchi type $I$ model.

\emph{Secondly}, in \cite{ColIbaHoo97}, it has been shown that all the Bianchi models with an exponential potential $V=e^{k\phi}$ (except the contracting Bianchi type $IX$ model), isotropise at late times when $k^2<2$. If $k=0$, these models tend toward a De Sitter model and toward $t^{2k^{-2}}$ otherwise. If $k^2>2$, the Bianchi type $I$, $V$, $VII$ and $IX$ models might isotropise at late times. In the present paper, the form of the coupling constant corresponding to the theory studied in \cite{ColIbaHoo97} is $\sqrt{3+2\omega}\phi^{-1}=\sqrt{2}$. What can we deduce from our results? If we introduce these forms of $\omega$ and $U$ in the expression of $\ell$, we see that the necessary condition for the isotropisation of the Bianchi type $I$ model will be $k^2<6$. Then, the Universe is of De Sitter type if $k=0$. In the other cases, the metric functions behave as $t^{2k^{-2}}$. The inflation arises when $k^2<2$. All these results are in accordance with these of \cite{ColIbaHoo97} and \cite{KitMae92}. However, some differences exist which are not in contradiction with the previous quoted papers: we have shown that Universe might isotropise and is inflationary when $k^2<2$ but isotropisation is impossible if $k^2>6$. Between these two values, the necessary condition for isotropy is respected but no inflation can occur.

\emph{Last}, these results are agreed with these found in \cite{Fay00C}. In this paper where the Hyperextended Scalar Tensor theory with a potential is studied for the Bianchi type I model, it is shown that the Universe isotropises when $\int G e^{3\Omega}dt$ tends toward a constant, $G$ being the gravitational coupling function. If in this last expression we choose $G=1$ and $e^{3\Omega}\rightarrow t^{-3l^{-2}}$, we find that isotropisation arises if $l^2<3$, in agreement with the above results.

Lets say few words about the power law potential, $U=\phi^{k}$. We can show from the asymptotical equation for $\phi$ defined above, that when $\Omega\rightarrow -\infty$, $\phi\rightarrow +\infty$ if $k<0$ (if $k>0$, the scalar field is not real). Then $\ell\rightarrow 0$ and isotropisation systematically leads to an asymptotical De Sitter model.\\

The result of this paper is not only a necessary condition for the isotropisation of the Universe. We have also derived the asymptotical behaviour of the metric functions and thus some conditions for a late time inflationary behaviour. It is a strong theoretical constraint on the forms of $\omega$ and $U$ so that the Universe be physically realistic at late times if we consider that it can be described by a Generalized Scalar-Tensor theory with a massive scalar field in the Bianchi type $I$ model. We have checked the compatibility of our results with these of the important No Hair Theorem and these of Kitada et al and Cooley et al that are here unified in a single condition. To our knowledge, there is no paper mixing the Hamiltonian technique and the dynamical systems theory with so many arbitrary functions. It seems to be a fruitful method in the case studied here mainly because it allows to calculate the equilibrium points as function of the potential and Brans-Dicke function. Then from mathematical constraints on the equilibrium points, we deduce constraints for these undetermined quantities. In future papers, we will see that we get the same type of results when we introduce a perfect fluid and we will extend this method to more general theory such has Hyperextended Scalar Tensor ones or other Bianchi models. 
%------------------------------------------------------------------------------------------------------------------------------------------------------------------------------------------%
\section{Appendix}\label{A1}
\underline{Equation for x}\\
From (\ref{9}) and (\ref{10a}), we deduce:
%-------------------------EQUATION-------------------------------------%
\begin{equation}\label{a1}
\dot{x}=3R^2y^2x
\end{equation}
\underline{Equation for y}\\
By deriving (\ref{10b}), we find:
%-------------------------EQUATION-------------------------------------%
\begin{equation} \label{a2}
\dot{\phi}=2L_2^{-1}x^{-2}y(\dot{y}-3R^2y^3)
\end{equation}
with $L_2=(e^{-6\Omega}U)_\phi$. By using (\ref{6}) and (\ref{10c}) to express $p_\phi$, we find:
%-------------------------EQUATION-------------------------------------%
\begin{equation} \label{a3}
\dot{\phi}=12zL_3^{-1}
\end{equation}
with $L_3=(3+2\omega)^{1/2}\phi^{-1}$
Consequently, (\ref{a2}) and (\ref{a3}) gives:
%-------------------------EQUATION-------------------------------------%
\begin{equation} \label{a4}
\dot{y}=6L_2L_3^{-1}x^2y^{-1}z+3R^2y^3
\end{equation}
\underline{Equation for z}\\
by using (\ref{10c}) to express $p_\phi$ and by deriving this expression, we get:
%-------------------------EQUATION-------------------------------------%
\begin{equation} \label{a5}
\dot{p_\phi}=L_3x^{-1}\dot{z}-3R^2L_3x^{-1}y^2z-12\phi^{-1}x^{-1}z^2+12L_4L_3^{-2}\phi^{-2}x^{-1}z^2
\end{equation}
with $L_4=\omega_\phi$. From (\ref{8}) and by using the fact that $U^{-1}U_\phi=L_2x^2y^{-2}+1/2L_3z^{-1}$, it comes:
%-------------------------EQUATION-------------------------------------%
\begin{equation} \label{a6}
\dot{p_\phi}=-12\phi^{-1}x^{-1}z^2+12L_4L_3^{-2}\phi^{-2}x^{-1}z^2-\frac{R^2}{2}L_2x-\frac{R^2}{4}L_3x^{-1}y^2z^{-1}
\end{equation}
From the equations (\ref{a5}) and (\ref{a6}), we derive:
%-------------------------EQUATION-------------------------------------%
\begin{equation} \label{a7}
\dot{z}=3R^2y^2z-\frac{R^2}{2}L_2L_3^{-1}x^2-\frac{R^2}{4}y^2z^{-1}
\end{equation}
Before getting the equation (\ref{12}) to (\ref{14}), we have to evaluate the term $L_2L_3^{-1}x^2y^{-2}$. After few calculations, we find $-(2z)^{-1}+\phi U^{-1}U_\phi(3+2\omega)^{-1/2}$.


\section*{References}
\begin{thebibliography}{10}

\bibitem{Per99}
S.~Perlmutter et~al.
\newblock Measurements of ${\Omega}$ and ${\Lambda}$ from 42 {H}ight-{R}edshift
  {S}upernovae.
\newblock {\em Astrophysical Journal}, 517.

\bibitem{Rie98}
Riess et~al.
\newblock Observational evidence from {S}upernovae for an accelerating
  {U}niverse and a cosmological constant.
\newblock {\em Astrophysical Journal}, 116:1009, 1998.

\bibitem{ChiJakPav00}
Luis~P. Chimento, Alejandro~S. Jakubi, and Diego Pav{\'o}n.
\newblock Enlarged quintessence cosmology.
\newblock {\em gr-qc/0005070.}, 2000.

\bibitem{Lan00}
A.~E.~Lange et~al.
\newblock First estimation of cosmological parameters from {B}oomerang.
\newblock astro-ph/0005004, 2000.

\bibitem{BraDic61}
C.Brans and Robert~H. Dicke.
\newblock Mach's principle and a relativistic theory of gravitation.
\newblock {\em Phys. Rev.}, 124, 3:925--935, 1961.

\bibitem{ColHaw73}
C.~B. Collins and S.~W. Hawking.
\newblock Why is the universe iostropic.
\newblock {\em Astrophys. J.}, 180:317--334, 1973.

\bibitem{Nar72}
Hidekazu Nariai.
\newblock Hamiltonian approach to the dynamics of {E}xpanding {H}omogeneous
  {U}niverse in the {B}rans-{D}icke cosmology.
\newblock {\em Prog. of Theo. Phys.}, 47,6:1824, 1972.

\bibitem{Mis62}
C.~W. Misner.
\newblock {\em Phys. Rev.}, 125:2163, 1962.

\bibitem{WaiEll97}
J.~Wainwright and G.F.R. Ellis, editors.
\newblock {\em Dynamical Systems in Cosmology}.
\newblock Cambridge University Press, 1997.

\bibitem{Fay00A}
S.~Fay.
\newblock Hamiltonian study of the generalized scalar-tensor theory with
  potential in a {B}ianchi type {I} model.
\newblock {\em Class. Quantum Grav.}, 17:891--902, 2000.

\bibitem{MatRyaTot73}
R.~A. Matzner, M.~P. Ryan, and E.~T. Toton.
\newblock The {B}rans-{D}icke theory and anisotropic cosmologies.
\newblock {\em Nuovo Cim.}, 14B:161, 1973.

\bibitem{ColIbaHoo97}
A.A.Coley, J.~Ib{\`a}{\~n}ez, and R.J. van~den Hoogen.
\newblock {\em J. Math. Phys.}, 38:5256, 1997.

\bibitem{KitMae92}
Y.~Kitada and M.~Maeda.
\newblock {\em Phys. Rev.}, D45:1416, 1992.

\bibitem{Fay00C}
S.~Fay.
\newblock Exact solutions of the {H}yperextended {S}calar {T}ensor theory with
  potential in the {B}ianchi type {I} model.
\newblock {\em Class. Quantum Grav.}, 18:45, 2001.
\end{thebibliography}
\end{document}